\documentclass[sigconf]{acmart}
\renewcommand\footnotetextcopyrightpermission[1]{}
\usepackage{booktabs} % For formal tables
\usepackage{mathtools}

\setcopyright{none}
\begin{document}
\title{Fiducia: A Personalized Food Recommender System for Zomato}
%\subtitle{Extended Abstract}

\author{Mansi Goel, Ayush Agarwal, Deepak Thukral, Tanmoy Chakraborty}
\affiliation{\institution{IIIT-Delhi, India}}
\email{[mansi14062, ayush14029, deepak14036, tanmoy]@iiitd.ac.in }

\if 0
\author{Ayush Agarwal}
\affiliation{\institution{IIIT-Delhi, India}}
\email{ayush14029@iiitd.ac.in}

\author{Deepak Thukral}
\affiliation{\institution{IIIT-Delhi, India}}
\email{deepak14036@iiitd.ac.in}

\author{Tanmoy Chakraborty}
\affiliation{\institution{IIIT-Delhi, India}}
\email{tanmoy@iiitd.ac.in}
\fi

\begin{abstract}
%With the advent of systems that can recognize the opinion of people and provide suggestions based on them, it was only a matter of time before the concept was utilized in the food interest. The extensive amount of review data which is available on food search and discovery service websites can be leveraged to provide restaurant insights. 
This paper presents {\em Fiducia}, a food review system involving a pipeline which processes restaurant-related reviews obtained from Zomato (India's largest restaurant search and discovery service). Fiducia is specific to popular caf\'e food items and manages to identify relevant information pertaining to each item separately in the reviews. It uses a sentiment check on these pieces of text and accordingly suggests an appropriate restaurant for the particular item depending on user-item and item-item similarity. Experimental results show that the sentiment analyzer module of Fiducia achieves an accuracy of over 85\% and our final recommender system achieves an RMSE of about $1.01$ beating other baselines.
\end{abstract}

\maketitle

	\section{Introduction}
Zomato\footnote{\url{https://www.zomato.com/india}} is the largest Indian restaurant directory application used to suggest nearby restaurants according to the favored cuisines, prices and other amenities like home delivery. 
%We believe that Zomato is a great platform to find suitable restaurants for users. 
It hosts vast information about each restaurant which is not clustered properly according to the users' convenience. If we manage to get a collection of restaurants recognized according to user preferences, the usability of the application will greatly increase. Zomato fails to provide details at the individual item level. A restaurant can be very popular but it might not be famous for a certain food item; whereas a less popular restaurant might excel in that. Usually customers need to go through the reviews to find out what's the best the restaurant serves and what should be avoided. \\
{\bf Problem Definition:}
In this paper, we aim to design a system that maps customers' choice of food to the place where they can be availed the best. The input of our system will be in the form of a preference food item of the user, and the recommendation generated will be based on restaurants which have a reviewed acclaim of that item. We also provide a peek into some of the side food items which go along well with the user's choice of items. \\
%
%Sentiment Analysis and recommendation systems is a growing area of machine learning and natural language processing with extensive research ranging from document level classification to personalized menu recommendation systems. Current endeavours in sentiment checks and recommendation set ups are dominated by supervised learning in the form of SVMs and SVDs and unsupervised learning has been explored in the form of clusterings and graph projections (Sawant et. al. 2013).  
{\bf System Description and State-of-the-art:} Fiducia leverages complex and fine-tuned structures in the form of a Naive Bayes Classifier and LSTMs. Previous work in recommendation of the Yelp food review dataset has relied on techniques like clustering and graph projections, matrix decompositions and SVM Ranking models. Collaborative filtering (CF) approaches on the dataset have been explored in \cite{nikulin2014hybrid} while work in user preference based recommendation has been addressed in \cite{zeng2016restaurant}. We do not rely on structured databases of recipes, in which a complete knowledge is available, but deal with fragmented information extracted solely from the user reviews of restaurants. Moreover, Fiducia deals with the problem of menu recommendation not in an abstract way, as a standalone problem, but contextualized to the user environment. Additionally, we also use techniques for topic modeling and community detection to determine suitable side dishes for user preference dishes.

To our knowledge, {\em this is the first work to recommend restaurants based on fragmented individual food item reviews on Zomato.}
\vspace{-0.3cm}
\begin{figure}[h]
	\centering
	\includegraphics[scale=0.22]{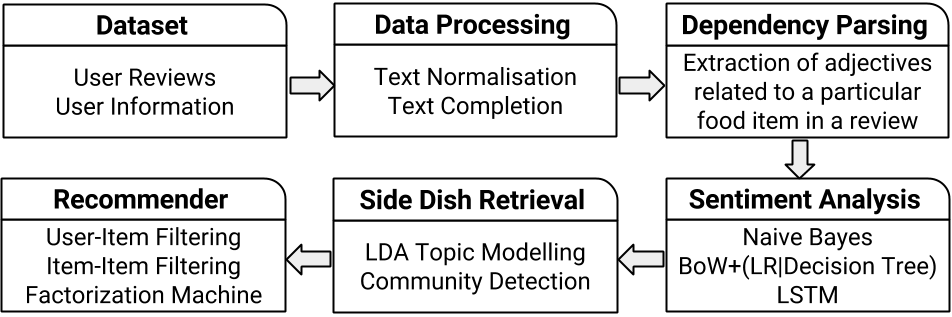}
	\vspace{-3mm}
	\caption{Flow diagram of Fiducia system. }\label{workflow}
\end{figure}
\vspace{-0.5cm}

\section{System Architecture}
Figure \ref{workflow} shows the overall architecture of our system.\\
\textbf{(i) Dataset}: The Zomato API provides exhaustive information about each food joint by different filters like location, cuisine etc.
However, it limits the number of reviews returned from every restaurant to 5. 
%We used the Selenium Module in Python using Firefox WebDriver to parse the data for 
We collected 100 popular restaurant profiles including  
Name, Rating, Cuisine, Reviews along with the rating and reviewer identity of each review.
%Data acquired through scraping contained unicode characters which needed to be removed.
Reviews on Zomato are written by common users mostly in code-mixed language (a mix of Hindi and English). We translated all such occurrences using Google Translate API to complete English text. Basic pre-processing steps included removing stop-words, assigning emoticons to its sentiment classes, replacing slangs with their intended meaning etc.
\newline
\textbf{(ii) Dependency Parsing}: Data extracted through Zomato contains a combined sentiment for multiple food items embedded in the same review. Object-wise sentiment detection problem is crucial for the recommender as it provides the user preferences for individual items especially in the cases of different sentiments for items expressed in the same review which would not be captured by the overall review sentiment. We use Stanford Dependency Parser\cite{chen2014fast} to identify the words intended for each item in a review. We create a vector of these fragments and use a sentiment analyzer framework on top of it to categorize the sentiment of each item. 
\newline
\textbf{(iii) Sentiment Analysis:}  We classify a fragment related to an item using three types of classifiers: (a) Naive Bayes Classifier gives us the probability of positive and negative sentiment for the item.
%5-fold cross validation of the classifier showed that more training data could increase the performance while increasing number of features didn't have any significant impact. The data had the problem of high variance with multiple sentiments expressed in the same review. Thus, more data was collected for a better classification.
%
(b) Bag of Words (BoW) model is used on each review as a document to make a dictionary containing the frequency of the vocabulary which is being built as we fit the data in the vectorizer. Through this we tokenize and count the word occurrences of a minimalistic corpus of reviews. Using the vocabulary we created a feature vector matrix of 0s and 1s according to the availability of a vocabulary element in that review. We tried training this vector with two classifiers - using Logistic Regression and using Decision Tree. (c) We also use Long Short Term Memory (LSTM). Simple RNNs are infamous for the vanishing gradient problem during backpropagation. This rises in cases of text snippets which are of longer length (similar to our case where the reviews, even for a single dish could be quite long). LSTMs  preserve information for a long time using a circuit that implements an analog memory cell.
Each review is mapped to a real vector domain in the embedding process which is the input to the network and the output consists of a number between $-1$ (neg. class) and $1$ (pos. class) which denotes the sentiment. 
\newline
\textbf{(iv) Main Food Item Recommendation}: We use two approaches in memory-based collaborative filtering -- User-Item Filtering (Model 1) and Item-Item Filtering (Model 2). A user-item filtering takes a particular user, finds users that are similar to that user based on similarity of ratings as $sim_u(u_m, u_n)$ = $ \frac {(u_i . u_j)} {( |u_i| |u_j| )}$ and recommends items that those similar users liked. We use cosine similarity as the distance metric. Item-item filtering takes an item, finds users who liked that item, and searches for other items that those users or similar users also liked as $sim_i(i_m, i_n)$ = $\frac {(i_m . i_n)} {( |i_m| |i_n| )}$. To make the final prediction (Eqs. $1$ and $2$), we use similarity as a weight, and normalize it with similarity of the ratings to stay within $-1$ and $1$.
\vspace{-0.3mm}
\begin{equation}\small
\text{Prediction for Model 1: } \hat{x}_{k,m} = \overline{x}_k + \frac {\sum\limits_{u_a} sim_u(u_k, u_a) ( x_{a,m} - \overline{x}_{u_a} )} {\sum\limits_{u_a} |sim_u(u_k, u_a)|} 
\end{equation}
%\vspace{-3mm}
\begin{equation}\small
\text{Prediction for Model 2: }
\hat{x}_{k,m} = \frac {\sum\limits_{i_b} sim_i(i_m, i_b) ( x_{k,b} )} {\sum\limits_{i_b} |sim_i(i_m, i_b)|}
\end{equation}
where $x_{a,m}$ represents the rating of item m of a restaurant by a user \textit{a}, $\overline{x}_k$ is mean of all ratings provided by user k and $\overline{x}_{u_a}$ is the mean of ratings provided to that item. 
We further use Factorization Machine which uses stochastic gradient descent with adaptive regularization as a learning method. It adapts the regularization automatically while training the model parameters. 
%It had superior performance. It works well with sparse data and ensures that there is no overfitting. 
We use a learning rate of 0.001 with 100 iterations, and task regression in the model. 
\newline
\textbf{(v) Side Food Item Recommendation}: Sometimes customers prefer to visit a restaurant where the sides frequented with the food items are equally good. An example of this can be observed in some reviews where the lack of a satisfactory serving of (say,) garlic bread with pasta is criticized. We aim to check the availability and quality of the side by employing two simple associative techniques.

\textbf{Topic Modeling:}
Taking each review as a document, we collate them for every restaurant. We use Latent Dirchilet Allocation to train our model for constructed corpus of each restaurant. This way, top 10 topics are identified with each topic containing 10 words from which we identify the items separately. It not only provides us with the most common items being talked about in the reviews but also provides us with the possibility of a group of items preferred together if they appear in the same topic.

\textbf{Community Detection:}
The concept of communities is applied to the graph of items appearing in reviews. Taking the items as nodes and edges according to their mentions in the reviews, a graph is constructed.
On this graph, we then apply Louvain algorithm \cite{louvain,Chakraborty:2017} for static community detection and obtain items appearing frequently together.
Among $51$ dishes we have in total, we find 6 distinct communities. 
\newline
We use this output as an additional feature in the recommender to evaluate the restaurant with respect to one of the primary items.

\section{Evaluation and Results}
\noindent\textbf{Evaluation of Sentiment Analyzer:}
After preprocessing, we had a total $3,131$ reviews in our dataset. For the evaluation of the sentiment analyzers, we keep 80\% (2504 samples) for training, and 20\% (627 samples) for testing.
We use F-score as our evaluation metric.
\if{0}
\begin{equation}
F1 score=\frac{2*TP}{2*TP+FP+FN}
\end{equation}
where $TP$ is the number of instances which are correctly
predicted, $FP$ is the number of instances wrongly classified as positive and $FN$ is the number of instances wrongly classified
negative.
\fi
\if{0}
\begin{equation}
RMSE = \frac {\sum\limits_{} (x_i - \hat{x}_i)^2} {N} 
\end{equation}
\fi
If we rely on using thresholding on restaurant rating (scale of 1-5) provided by Zomato as ground-truth, in case of low-rated (<2.0) restaurants we bias the evaluation with negative words, while the opposite happens in case of high-rated (>3.0) restaurants. Therefore, we further manually annotated all $3,131$ reviews (annotations were done by $3$ annotators with an inter-annotator agreement of $0.82$). Table \ref{Table 1} shows that LSTM outperforms others across all the ground-truth datasets. Best performance ($0.90$ F-score) is obtained on manually-annotated ground-truth dataset.

\begin{table}[!t]
	% \centering
	\scalebox{0.85}{
		\begin{tabular}{|c|c|c|c|c|  }
			\hline
			Data Set & Naive & BoW  & BoW +& LSTM \\
			& Bayes &+LR &DT & \\
			\hline
			Rated 2.0 & 0.77 & 0.81 &  0.86 & 0.81\\
			Rated 2.5 & 0.76 &  0.82 &  0.82 & 0.84 \\
			Rated 3.0 &  0.74 & 0.79 & 0.81 & 0.83 \\
			Manual & \textbf{0.80} & \textbf{0.84}& \textbf{0.86} & \textbf{0.90} \\
			% Annotated& & & & \\
			\hline
	\end{tabular}}
	\caption{F-score of the sentiment analyzers across multiple ground-truth datasets.}
	\label{Table 1}
	\vspace{-8mm}
\end{table}

\begin{table}[!t]
	\centering
	\scalebox{0.9}{
		\begin{tabular}{|c|c|c|c|}
			\hline
			CF Method & RMSE & MAE & Precision\\
			\hline
			Baseline & 1.952&1.071& 0.48\\
			User-Item Filter & 1.054&0.634&0.72\\
			Item-Item Filter& 1.059&0.661&0.70\\
			Factorization Machine & {\bf 1.010}&{\bf 0.609}&{\bf 0.74}\\
			
			\hline
	\end{tabular}}
	\caption{Results of the recommender system.}
	\label{Table2}
	\vspace{-9mm}
\end{table}

\noindent\textbf{Evaluation of Recommender System:} To evaluate our recommendation system, we use Root Mean Squared Error (RMSE), Mean Absolute Error (MAE) and precision. Depending on a person's similarity with another user or the items in the reviews being similar, we make a recommendation for a restaurant w.r.t the person, and then evaluate it with the actual rating of the item in the person's review. The baseline prediction is made by simply suggesting the restaurant with the highest number of positive reviews. Table \ref{Table2} shows that Factorization Machine outperforms others with a precision of $0.74$.

\section{Conclusions}
We designed Fiducia, a recommender system which suggests restaurants according to user's food item preference based on the reviews received by Zomato about these restaurants. We also kept in mind the side items favored while making the recommendation. We believe that this information may be useful for the extensive review data available on such sites. 
%We will make the data, annotation and code publicly available upon acceptance of the paper.  
%Our study could be extended to cover all cuisines and restaurants available. Collecting data from user profiles and the idea of exploiting social ties between reviewers might also be prudent to be explored further.
%\end{document}  % This is where a 'short' article might terminate

\bibliographystyle{ACM-Reference-Format}

%\bibliography{sample-bibliography} 

\end{document}